\documentclass[twocolumn,english]{article}
\usepackage[T1]{fontenc}
\usepackage{amssymb}
\usepackage{graphicx}

\makeatletter
\usepackage{a4}
\input{epsfig.sty}
\def\fnum@table{\tablename~{\bf\thetable}}
\def\fnum@figure{\figurename~{\bf\thefigure}}
\def\tablename{\footnotesize{\bf Table}}
\def\figurename{\footnotesize{\bf Figure}}

\def\be{\begin{equation}}
\def\ee{\end{equation}}

\makeatother

\usepackage{babel}
\begin{document}

\title{\textbf{On the model uncertainties for the predicted maximum depth
 of extensive air showers}}

\author{Sergey Ostapchenko  and G\"unter Sigl\\
\textit{\small Universit\"at Hamburg, II Institut f\"ur Theoretische
Physik, 22761 Hamburg, Germany}\\
}

\maketitle
\begin{center}
\textbf{Abstract}
\par\end{center}
A quantitative analysis of model uncertainties
for calculations of the 
maximum depth 
 of proton-initiated extensive air showers (EAS) has been performed. Staying within the standard physics picture and using the conventional approach to the treatment of high energy  interactions, we found that present uncertainties on the energy dependence of the inelastic cross section, the rate of diffraction, and the inelasticity of hadronic collisions allow one to increase the predicted
 average EAS maximum depth 
  by about 10 g/cm$^2$. Invoking more exotic assumptions regarding a potentially significant modification of the parton hadronization procedure by hypothetical collective effects, we were able to change drastically the predicted energy dependence of the inelasticity of proton-air interactions and to increase thereby the predicted EAS maximum depth by up to $\simeq 30$   g/cm$^2$. However, those latter modifications are disfavored by the data of the LHCf experiment, regarding forward neutron production in proton-proton collisions at the Large Hadron Collider,
and by measurements of the  muon production depth by the Pierre Auger Observatory.

\section{Introduction\label{intro.sec}}
One of the exciting directions of astroparticle physics research is connected to
experimental studies of ultra-high energy cosmic rays (UHECRs). Because of the
extremely low incoming flux of UHECRs, one is forced to employ indirect
detection methods: inferring the properties of the primary particles from
measured characteristics of  extensive air showers (EAS) --
nuclear-electromagnetic cascades initiated by interactions of UHECRs in the
atmosphere \cite{nag00}. Consequently, the success of such studies depends
on the quality of the                           description of EAS development by the corresponding numerical
tools, like the CORSIKA program \cite{hec98}. A special role in such EAS
simulation procedures is played by Monte Carlo (MC) generators of high energy
hadronic interactions, used to describe the backbone of an air shower -- the
cascade of interactions in the atmosphere of both primary cosmic ray (CR) 
particles and of secondary hadrons produced  \cite{eng11}.

All of the above fully applies to investigations of UHECR composition, based on
studies of the longitudinal extensive air shower development, notably, on
measurements of EAS maximum depth, $X_{\max}$,
 corresponding to the maximum of the 
charged particle profile of an air shower. Modern fluorescence detectors allow
one to measure $X_{\max}$ for individual events rather accurately 
(see \cite{kam12} for a review).
On the other hand, predictions of EAS simulation procedures for
the average EAS maximum depth,  $\langle X_{\max}\rangle$,
being driven by the treatment of primary particle interactions with air nuclei,
are rather seriously constrained by experimental data from the Large Hadron 
Collider (LHC) (see, e.g.\ \cite{ost19}). Therefore, somewhat surprising
is a certain tension between the predictions of air shower simulations and the
experimental data of the Pierre Auger Observatory, 
regarding distributions of the
shower maximum depth 
 \cite{abr13,PierreAuger:2024neu}. 
More specifically, Auger data demonstrate that both  the elongation rate
${\rm ER}(E_0)=d\langle X_{\max}(E_0)\rangle /d\lg E_0$ and the width of  $X_{\max}$ distributions,
  $\sigma (X_{\max})$, decrease with the primary energy $E_0$ faster than predicted
  by EAS simulations corresponding to an energy-independent UHECR composition,
  thereby indicating a change towards heavier CR primaries at ultra-high energies
  \cite{pao14,pao14a}. However, the observed energy dependence of  $\sigma (X_{\max})$
  implies a faster change of the composition, compared to the one deduced from
   ${\rm ER}(E_0)$, e.g., if one employs the QGSJET-II-04 model \cite{ost11,ost13} for
   interpreting the measurements. As demonstrated in \cite{PierreAuger:2024neu},
   a consistent interpretation of the data requires a significantly slower extensive
   air shower development, compared to current EAS simulation results. Consequently,
of significant importance is to quantify the range of uncertainty for  $X_{\max}$
predictions, related to potential modifications of the treatment of high energy
hadronic interactions, allowed by LHC data.

Such a quantitative investigation is the subject of the current work. We consider
all plausible changes of the interaction treatment, in the framework of the 
QGSJET-III model \cite{ost24,ost24a}, which may potentially improve the agreement
of the model predictions with Auger data, while staying within the standard
physics picture. Like in our previous study of model uncertainties for the 
predicted EAS muon content \cite{nmu24}, our investigation  is guided by three
basic principles:  i) the changes of the corresponding modeling are performed at a 
microscopic level; ii) the considered modifications are restricted by the requirement
not to contradict basic physics principles; iii) the consequences of such changes,
regarding a potential (dis)agreement with relevant accelerator data, are analyzed.

The outline of the paper is as follows. In Section \ref{inter.sec}, we identify
basic characteristics of hadronic interactions, which are of direct relevance to
 $X_{\max}$ predictions. In Section \ref{modif.sec}, we consider various 
 modifications of the treatment of high energy interactions, which change such
 characteristics, and study the corresponding consequences both for the predicted EAS maximum depth 
 and for comparisons with relevant LHC data. We summarize our 
 investigation in Section \ref{summary.sec}.
 
\section{Characteristics of high energy interactions, relevant for  $X_{\max}$
predictions \label{inter.sec}}

Unlike the EAS muon content depending on the whole history of 
the nuclear cascade in the atmosphere, the maximum depth of a proton-initiated 
extensive air shower is largely governed by an interaction of the primary particle.
As demonstrated, e.g., in \cite{nmu24}, modifying the treatment of secondary
pion-air collisions has a rather weak impact on the calculated 
$\langle X_{\max}\rangle$.
Therefore, predictions of EAS simulations for the longitudinal extensive air shower
development are rather seriously constrained by available LHC data on proton-proton
and, to some extent, proton-nucleus interactions.

Of primary importance for  $\langle X_{\max}\rangle$
 predictions is the proton-air inelastic
cross section $\sigma^{\rm inel}_{p-{\rm air}}$ since it controls the mean free
path  of protons in air, 
 $\lambda_p = m_{\rm air}/\sigma^{\rm inel}_{p-{\rm air}}$,
$m_{\rm air}$ being the average mass of air nuclei, and thereby the starting
point for a nuclear cascade. The energy rise of  $\sigma^{\rm inel}_{p-{\rm air}}$
is the main factor which causes the decrease of the elongation rate ${\rm ER}(E_0)$
 with increasing  primary energy $E_0$.
The inelastic proton-proton cross section  $\sigma^{\rm inel}_{pp}$
has been measured at LHC with a high precision; the difference between the
results of the TOTEM and ATLAS experiments, based on   Roman Pot techniques,
is  $\simeq 2.6$\%   at the center-of-mass (c.m.)
 energy $\sqrt{s}=13$ TeV \cite{ant19,aad23}.
Using the corresponding values for calculating  $\sigma^{\rm inel}_{p-{\rm air}}$,
within the Glauber-Gribov formalism \cite{gla59,gri69}, one obtains even a smaller
difference for the  proton-air cross section since it is largely dominated
by the nuclear size.

Also of importance for calculations of EAS 
maximum depth is the treatment 
of inelastic diffraction. Diffractive interactions of primary protons are characterized
by a small inelasticity  $K^{\rm inel}_{p-{\rm air}}$, i.e., by a small relative
energy loss of ``leading'' (most energetic) secondary nucleons. This is especially
so for a diffractive excitation of a target nucleus, in which case the incoming proton
looses a tiny fraction of its initial energy, 
$K^{\rm inel}_{p-{\rm air}}\simeq M_X^2/s\ll 1$, $M_X$ being the diffractive state
mass, i.e., one   essentially deals with a quasi-elastic collision. 
Therefore, the main effect
of diffraction amounts to a ``renormalization'' of the inelastic proton-air cross 
section, merely subtracting its diffractive part, $\sigma^{\rm diffr}_{p-{\rm air}}$:
 $\sigma^{\rm inel}_{p-{\rm air}}\rightarrow \sigma^{\rm inel}_{p-{\rm air}}
-\sigma^{\rm diffr}_{p-{\rm air}}$, and of  proton mean free path:
\begin{equation}
\lambda_p\rightarrow \lambda_p (1+\sigma^{\rm diffr}_{p-{\rm air}}/
\sigma^{\rm inel}_{p-{\rm air}})\,.
\label{eq:mfp}
\end{equation}
In addition, the  inelastic diffraction is closely related to  the inelastic 
screening effect \cite{gri69}; a higher diffraction rate corresponds to a smaller
 $\sigma^{\rm inel}_{p-{\rm air}}$, for a given  $\sigma^{\rm inel}_{pp}$,
 and vice versa.
 
 Finally, calculations of  $\langle X_{\max}\rangle$
  depend on the predicted energy dependence 
 of the average inelasticity of proton-air interactions, which is governed by the
 treatment of nondiffractive collisions. A smaller  $K^{\rm inel}_{p-{\rm air}}$
 corresponds to a slower energy dissipation for leading nucleons, hence, to a larger
 fraction of the energy of the primary particle, being retained in the 
 hadronic cascade, and to a slower EAS 
 development.  An enhancement of
 secondary particle production, with increasing energy, due to the energy rise
 of the multiple scattering rate, inevitably leads to an increase of the 
 inelasticity. However, the speed of the energy rise of   
 $K^{\rm inel}_{p-{\rm air}}$ is rather weakly constrained by LHC data and is
  highly model-dependent \cite{ost16}, being currently the main 
 source of model uncertainties for  $\langle X_{\max}\rangle$ predictions.
 
Coming now to fluctuations of EAS maximum depth,  characterized by 
the standard deviation $\sigma (X_{\max})$,
those are rather insensitive to the average inelasticity, being dominated by
variations of the proton mean free path, hence, depending mostly on 
$\sigma^{\rm inel}_{p-{\rm air}}$. In addition, they are influenced somewhat
by the rate of inelastic diffraction, where the  effective
``renormalization'' of the inelastic proton-air cross section and the
inelastic screening effect work in the same direction, e.g., 
 both effects contribute to enlarging  $\sigma (X_{\max})$
in case of a  higher diffraction.
Overall, the uncertainties of predictions for  $\sigma (X_{\max})$
 are rather small, as noticed already in \cite{alo08}, notably, thanks to
 measurements of total, elastic, and diffractive $pp$ cross sections by the
 TOTEM and ATLAS experiments  at LHC 
  \cite{ant19,aad23,ant13b,ant13a,ant13c,aad14,aab16,ant13,aad22,olj20}. 
 For example,
 the current $\simeq 3$\% difference between the results of TOTEM and ATLAS
 for  $\sigma^{\rm inel}_{pp}$ at $\sqrt{s}=13$ TeV translates itself into
 $\simeq 1.5$ g/cm$^2$ variation of $\sigma (X_{\max})$ [cf.\ Eq.\ (\ref{eq:mfp})]. 
 On the other hand, the impact on the fluctuations of  $X_{\max}$
 of present uncertainties  for $\sigma^{\rm diffr}_{pp}$ is at the level
 of few g/cm$^2$ \cite{ost14}.
 
 Regarding average characteristics of extensive air showers initiated
 by primary nuclei, e.g.\ $\langle  X_{\max} \rangle$,
 those are well described by the superposition model,
 i.e., coincide with a good accuracy with the corresponding
  contributions of $A$ proton-induced EAS of $A$ times smaller energy,
  for a primary nucleus of mass number $A$ (e.g.\ \cite{eng92,kal93}).
  This follows from the 
  relation between the inelastic
  nucleus-air cross section $\sigma^{\rm inel}_{A-{\rm air}}$
  and the mean number of interacting 
  projectile nucleons $\langle \nu_A\rangle$, per inelastic 
  collision \cite{bia76}:
 \begin{equation}
\langle \nu_A\rangle =\frac{A\,\sigma^{\rm inel}_{p-{\rm air}}}
{\sigma^{\rm inel}_{A-{\rm air}}}\,,
\label{eq:nwound}
\end{equation}
and from the possibility to approximate forward production spectra
of an $A$-air collision, for given $\langle \nu_A\rangle$, by the
ones of proton-air interactions:
 \begin{equation}
\left.\frac{dN^X_{A-{\rm air}}(E_0,E)}{dE}\right|_{\langle \nu_A\rangle}
\rightarrow \langle \nu_A\rangle \,\frac{dN^X_{p-{\rm air}}(E_0/A,E)}{dE}\,,
\label{eq:supspec}
\end{equation}
for any secondary particle $X$. As a consequence of Eq.\ (\ref{eq:nwound}),
mean number of projectile nucleons interacting in a given depth interval,
for an air shower initiated by nucleus $A$, is the same as for a 
superposition of  $A$ proton-induced EAS of $A$ times smaller energy \cite{kal93}.

Yet the superposition model is invalid for fluctuations of EAS characteristics
\cite{eng92,kal93,kal89}, e.g., for $\sigma (X_{\max})$,
 which are dominated by variations of the impact
parameter of nucleus-air collisions, causing large fluctuations of the number
of interacting projectile nucleons. While those are well-defined  in the 
Glauber-Gribov approach, an additional contribution to  $\sigma (X_{\max})$
comes from a fragmentation of the spectator part of the projectile 
nucleus \cite{eng92,kal93}. For example, for an extensive air shower initiated
by a primary iron nucleus, one obtains a factor two difference between the
predicted values of $\sigma (X_{\max})$, when considering two extreme
assumptions: a full breakup of the nuclear spectator part into separate
nucleons or keeping all non-interacting nucleons together, as a single
secondary nucleus \cite{kal93}. Since experimental data on nuclear
fragmentation are available at fixed target energies only, this could have
constituted a serious source of uncertainty for predictions of $\sigma (X_{\max})$, for nucleus-induced EAS. 
However, the relative yields of various nuclear fragments are
energy-independent, above few GeV per nucleon \cite{fre87}, which allows
one to reliably calibrate the fragmentation procedures, based on fixed
target data, and to safely extrapolate them to UHECR energies.

As follows from the above discussion, the main model uncertainties for
$\langle X_{\max}\rangle$ and  $\sigma (X_{\max})$ stem 
from the treatment of proton-air
interactions, while  an extension to the case of nuclear primaries is
rather well defined. Therefore we concentrate in the following on a study of proton-induced extensive air showers.

\section{Modifications of the treatment of proton-air collisions, 
leading to a larger $\langle X_{\max}\rangle$ \label{modif.sec}}

\subsection{Smaller cross section and  higher diffraction rate\label{modsig.sec}} 

\begin{figure}[htb]
\centering
\includegraphics[height=6cm,width=0.49\textwidth]{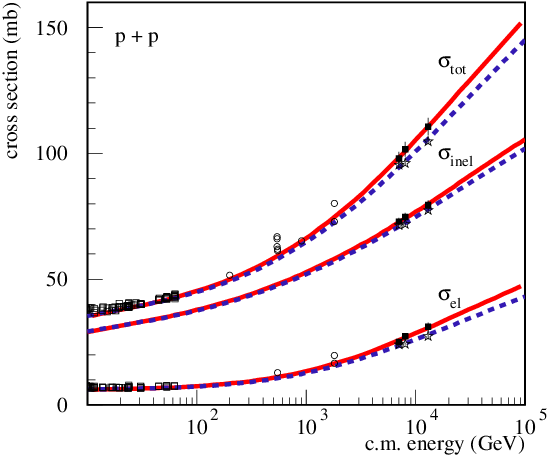}
\caption{C.m.\ energy dependence of the total, inelastic, and elastic $pp$ cross
sections,  calculated with  the default QGSJET-III model (red solid lines) 
 and with  the option characterized by an enhanced diffraction (blue dashed lines), 
compared to experimental data (points) from \cite{ant19,aad23,pdg}.}
\label{fig:sig-ht}       
\end{figure}

As discussed in Section \ref{inter.sec}, both a smaller proton-air cross section
and a higher diffraction rate lead to a larger $\langle X_{\max}\rangle$ predicted. Here we
are going to study a combined effect of both, increasing the rate of low mass
diffraction (LMD) in the QGSJET-III model. The LMD treatment in QGSJET-III is
based on the Good-Walker (GW) formalism \cite{goo60}: considering a hadron to
be represented by a superposition of a number of GW Fock states characterized
by different sizes and different (integrated) parton densities. Since such
states undergo different absorption during a scattering process, one generally
has transitions of the initial hadrons into various low mass excited states
(see, e.g.\ \cite{kho21}, for the corresponding discussion). Obviously, to enhance
the LMD rate, one thus needs to enlarge the difference between transverse sizes
of the GW states. Using a twice smaller value for the ratio of the squared radii
of the smallest and largest GW states of the proton (parameter $d_p$ in 
QGSJET-III \cite{ost24}), we obtain cross sections for single diffractive-like
(SD-like) events\footnote{As discussed in \cite{ost24a}, the cross sections for 
SD-like events include contributions of both diffractive and nondiffractive
collisions characterized by a given structure of final states, i.e., containing 
a rapidity gap of certain size, devoid of secondary hadrons.},
 for different intervals of diffractive state mass
$M_X$, listed in Table \ref{tab: SD-totem}.
\begin{table*}[t]
\begin{tabular*}{1\textwidth}{@{\extracolsep{\fill}}lllllll}
\hline 
$M_{X}$ range, GeV & $<3.4$ & $3.1-7.7$  &  $7.7-380$  & $380-1150$  &  $1150-3100$ &
  $3.1-3100$\tabularnewline
\hline 
\hline 
TOTEM \cite{ant13a,olj20} & $2.62\pm 2.17$ & $1.83\pm 0.35$ & $4.33 \pm 0.61$ & $2.10 \pm 0.49$
 & $2.84 \pm 0.40$ & $11.10 \pm 1.66$ \tabularnewline
QGSJET-III & 3.22 & 1.41  & 3.19  & 1.51  & 6.38 & 12.49 \tabularnewline
option SD$^+$ & 4.20 & 1.50  & 3.06  & 1.34  & 6.17 & 12.07\tabularnewline
\hline 
\end{tabular*}
\caption{Cross sections (in mb) of SD-like $pp$ collisions at $\sqrt{s}=7$ TeV, 
for different ranges of mass $M_{X}$ of diffractive states produced, 
  calculated with  the default QGSJET-III model
 and with  the option characterized by an enhanced diffraction, 
compared to TOTEM data.
\label{tab: SD-totem}}
\end{table*}
 It is easy to see that the LMD rate, for $M_X<3.4$ GeV,
is enhanced by $\simeq 30$\%, compared to the default QGSJET-III model, while being still
compatible with the observations of the TOTEM experiment \cite{ant13a}.
Since the considered modification gives rise to a strong enhancement of the
inelastic screening effect (see, e.g.\ \cite{ost19a}, for the corresponding discussion),
it leads also to a sizable reduction of the calculated total, elastic, and inelastic
 $pp$ cross sections plotted in Fig.\ \ref{fig:sig-ht}, which now become compatible with the
 data of the ATLAS experiment \cite{aad23,aad14,aab16}.
 \begin{figure}[htb]
\centering
\includegraphics[height=6.cm,width=0.49\textwidth]{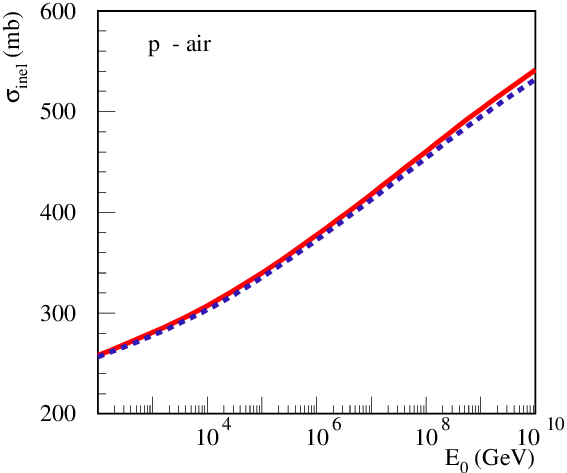}\hfill
\caption{Laboratory (lab.) energy dependence of   $\sigma^{\rm inel}_{p-{\rm air}}$,  
calculated  with the default QGSJET-III model
(red solid line) and  with  the option characterized by an enhanced diffraction 
 (blue dashed line).}
\label{fig:sigpair}       
\end{figure}%
 \begin{figure*}[htb]
\centering
\includegraphics[height=6.cm,width=0.49\textwidth]{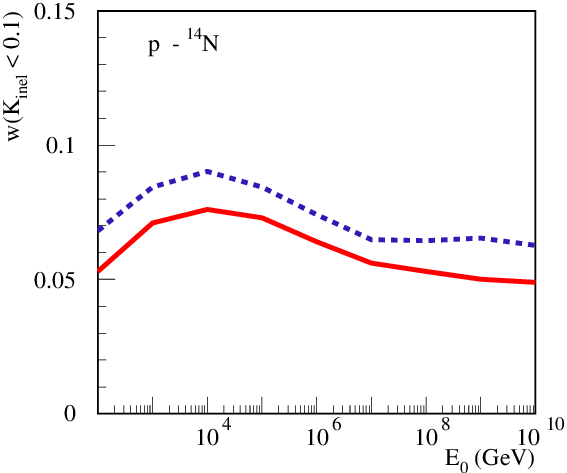}\hfill
\includegraphics[height=6.cm,width=0.49\textwidth]{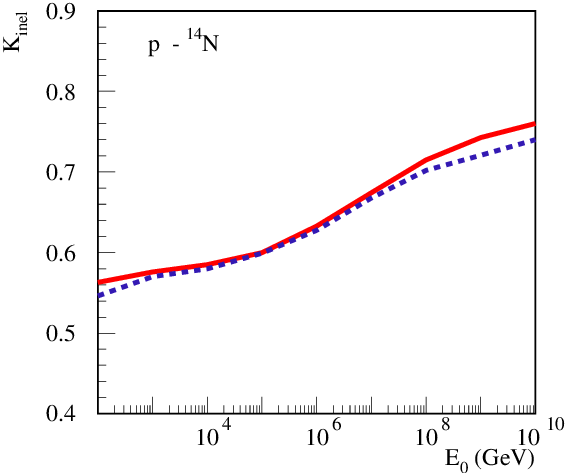}
\caption{Lab.\ energy dependence of  the probability of  diffractive-like interactions
(left) and  of  the inelasticity (right),
for  $p\,^{14}$N collisions, calculated with the default QGSJET-III model
(red solid lines) and   with  the option characterized by an 
 enhanced diffraction (blue dashed lines).
}
\label{fig:wdiffr}       
\end{figure*}%

 However, as anticipated in Section \ref{inter.sec}, the impact of such changes
 on the calculated inelastic proton-air cross section plotted in 
  Fig.\ \ref{fig:sigpair} is much more moderate: reaching only $\simeq 1$\% level at
  $E_0=10^{19}$ eV. On the other hand, one obtains a  significant enhancement of diffraction in proton-nucleus collisions. For example, for proton
  interactions with the most abundant air element, nitrogen, the increase
  of the rate of
  diffractive-like collisions characterized by a small inelasticity,
   $K^{\rm inel}_{p{\rm N}}<0.1$, reaches $\simeq 15$\% 
  level, as one can see in  Fig.\ \ref{fig:wdiffr} (left).
It is also noteworthy that such an enhancement of diffraction 
has a small impact on the average inelasticity of  proton-nitrogen  interactions
[cf.\ Fig.\ \ref{fig:wdiffr} (right)], the latter being dominated by the
treatment of nondifractive collisions.

\begin{figure*}[htb]
\centering
\includegraphics[height=6.cm,width=\textwidth]{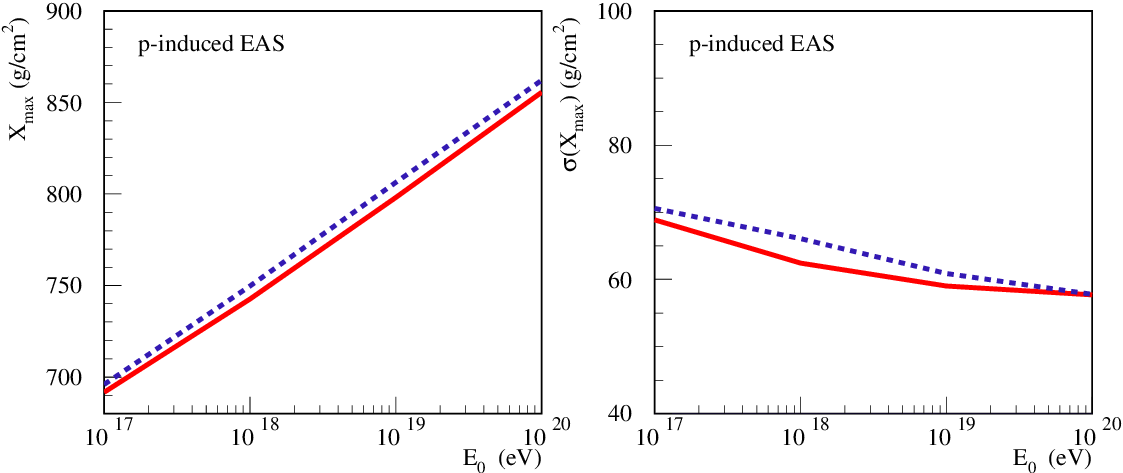}
\caption{Dependence on primary energy of  the average maximum depth 
(left) and of the corresponding standard deviation 
 (right),
for proton-initiated EAS, calculated with the default QGSJET-III model
(red solid lines) and    with  the option characterized by an 
 enhanced diffraction (blue dashed lines).}
\label{fig:xmax}       
\end{figure*}

As is obvious from the above-presented results, the changes of the 
predicted\footnote{Here and in the following, we perform EAS
 simulations, using the CONEX code \cite{ber07}.}
 $\langle X_{\max}\rangle$ and  $\sigma(X_{\max})$ plotted in  Fig.\ \ref{fig:xmax} are
driven by the increased rate of  diffractive-like proton-air collisions. 
The magnitude of these changes, up to $\simeq 8$ g/cm$^2$ for 
$\langle X_{\max}\rangle$ and
$\lesssim 4$  g/cm$^2$ for $\sigma(X_{\max})$, agrees well with the
corresponding results of \cite{ost14}.

\subsection{Slower energy rise of the inelasticity\label{modinel.sec}} 
Thus, the only possibility to predict a substantially larger EAS
maximum depth is to decrease the  inelasticity of nondiffractive
proton-air interactions. 
A violation of the Feynman scaling
in hadronic collisions is a well established experimental fact,
e.g., regarding  a steep 
 rise of   central pseudorapidity $\eta$ density of charged hadrons
produced in $pp$ interactions,
$\left.dN^{\rm ch}_{pp}/d\eta\right|_{\eta=0}$, with energy,
which stems from an increase
of multiple scattering rate. However, an approximate 
Feynman scaling of secondary hadron spectra at large values of
Feynman $x$ may still be allowed by experimental data.

Generally, the rate of multiple scattering in hadronic collisions
rises with energy, primarily, due to a fast increase 
of the rate of semihard processes leading to production of
hadron (mini)jets, 
which is related, in turn, to a steep low $x$ rise
of parton momentum distribution functions  of hadrons. Yet the rate of this 
increase may be tamed by nonlinear interaction effects, notably, regarding
a copious production of minijets of relatively small transverse momenta $p_t$.
On the other hand,  the inelasticity of high energy
collisions depends strongly on the choice of momentum distributions 
of  constituent partons for the  interacting hadrons (nuclei),
 involved in  numerous inelastic rescattering processes \cite{ost16,ost03,par11}.
  Choosing a softer  distribution,
$\propto x^{-\alpha}$, for  the  fraction $x$ of the initial light cone (LC) momentum of the parent  hadron, taken by such a parton, i.e., using a larger
value for  $\alpha$, one obtains a weaker  impact  of  multiple scattering
on forward hadron production, arriving to  an approximate 
Feynman scaling at large $x$ in the  $\alpha \rightarrow 1$ limit \cite{ost16}.

\begin{figure}[htb]
\centering
\includegraphics[height=6cm,width=0.49\textwidth]{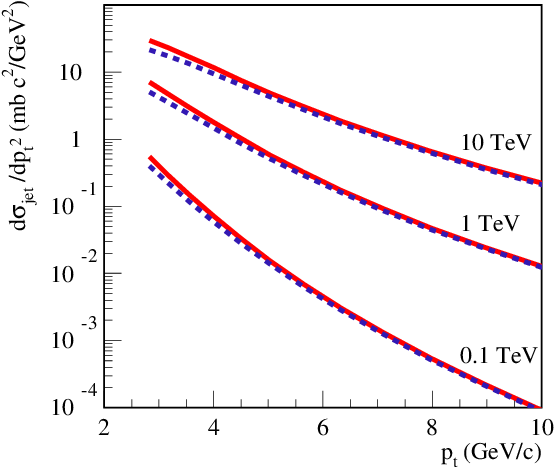}
\caption{Transverse momentum spectra for (mini)jet production in $pp$ collisions
at   $\sqrt{s}=10^2$, $10^3$, and  $10^4$ GeV, as indicated in the plot, 
as calculated with the default QGSJET-III model (red solid lines) and with 
the option characterized by twice stronger HT corrections (blue dashed lines).}
\label{fig:ptjet-ht}       
\end{figure}

\begin{figure*}[t]
\begin{centering}
\includegraphics[width=0.9\textwidth,height=6.5cm]{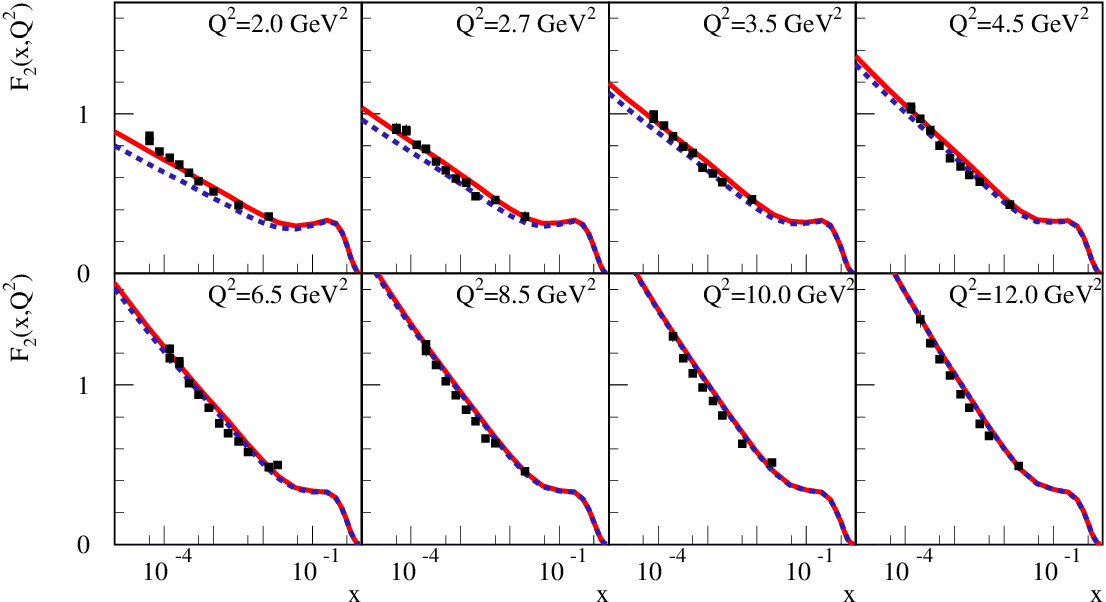}
\caption{$x$-dependence 
of the  proton  structure function $F_{2}(x,Q^{2})$, for different $Q^{2}$,
 as indicated in the plots, as calculated with the default QGSJET-III model (red solid
 lines) and with the option characterized by twice stronger HT corrections (blue dashed
 lines),  compared to HERA data \cite{aar10} (points).\label{fig:f2}}
\par\end{centering}
\end{figure*}

\begin{figure*}[htb]
\centering
\includegraphics[height=6.cm,width=\textwidth]{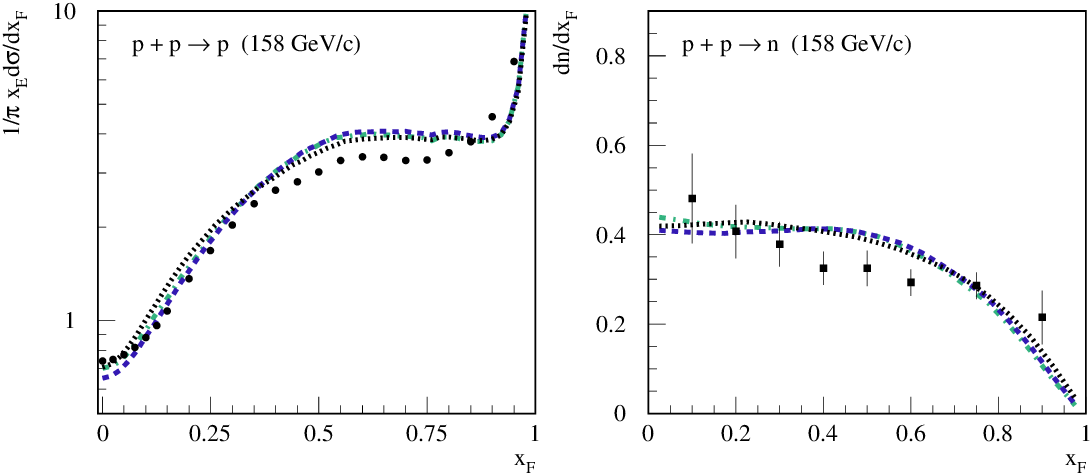}
\caption{$x_{\rm F}$-dependence of $p_t$-integrated invariant cross 
section 
for proton production (left)
and $x_{\rm F}$-distribution of neutrons (right) in c.m.\ frame,
for  $pp$  collisions at 158 GeV/c, compared to NA49 data \cite{ant10}.
Black dotted, blue dashed, and green dash-dotted lines correspond to calculations 
with the modified QGSJET-III model characterized by twice stronger 
HT corrections,  for the parameter  $\alpha_{\rm sea}=0.65$, 0.8, and 0.9, 
respectively.}
\label{fig:pp158}       
\end{figure*}

\begin{figure*}[htb]
\centering
\includegraphics[height=6.cm,width=0.49\textwidth]{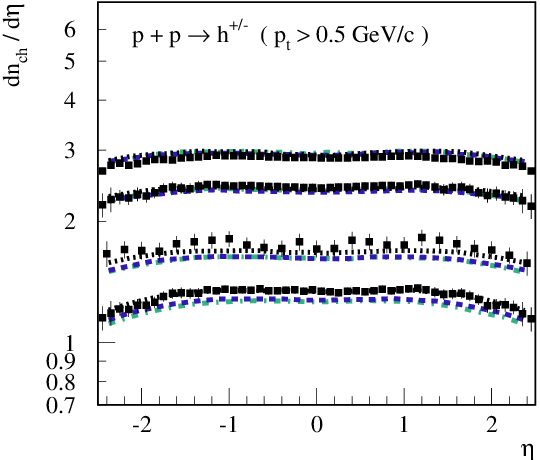}\hfill
\includegraphics[height=6.cm,width=0.49\textwidth]{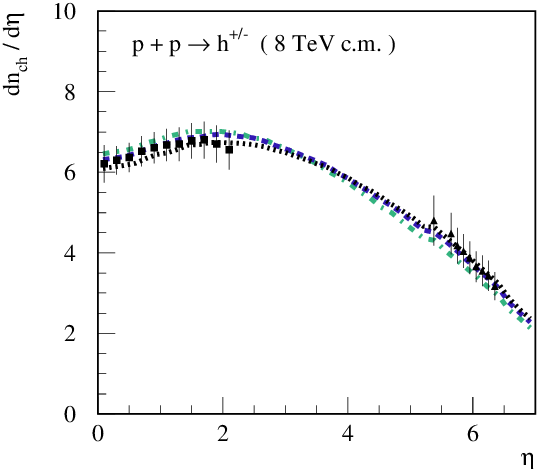}
\caption{Calculated pseudorapidity $\eta$ distributions of
charged hadrons in c.m.\ frame. Left: for $pp$ collisions at 
different $\sqrt{s}$ (from top to bottom: 13, 7, 2.36, and 0.9 TeV),
for hadron transverse momentum $p_t>0.5$  GeV/c, compared to ATLAS 
data \cite{aad11,aad16}. Right:  for $pp$ collisions at $\sqrt{s}=8$
TeV, compared to the data of CMS and TOTEM \cite{cha14}.
The notations for the lines are the same as in Fig.\   \ref{fig:pp158}.}
\label{fig:pp-lhc}       
\end{figure*}

To investigate the impact of a reduced multiple scattering rate,
 we increase the strength of higher
twist (HT) corrections to hard scattering processes in the QGSJET-III model,
choosing a twice larger value, $K_{\rm HT}=5$, for the corresponding
normalization parameter \cite{ost24}. As one can see in Fig.\ \ref{fig:ptjet-ht},
this way we decrease
significantly, by up to $\simeq 30$\%, the minijet production at small $p_t$.
It is noteworthy, however, that the considered change is an extreme one
since it causes a tension with HERA data on the low $x$ behavior, at small $Q^2$,
 of the proton structure function $F_2(x,Q^{2})$ plotted in Fig.\ \ref{fig:f2}.
 Additionally, we vary the parameter $\alpha_{\rm sea}$ which governs the
LC momentum distribution of constituent sea (anti)quarks in the model
($\propto x^{-\alpha_{\rm sea}}$) \cite{ost24a}: using 
$\alpha_{\rm sea}=0.8$ and $\alpha_{\rm sea}=0.9$, in addition to the
default value $\alpha_{\rm sea}=0.65$. For all the considered options,
we adjust other parameters of the hadronization procedure of the model
in order to keep a reasonable agreement with hadron production data,
both from fixed target experiments and from LHC, as illustrated in  
Figs.\ \ref{fig:pp158} and \ref{fig:pp-lhc}.
 Noteworthy is the modification of the $\eta$-dependence of the charged
 hadron yield, for increasing  $\alpha_{\rm sea}$ (cf.\ the plot on the
 right-hand side of Fig.\ \ref{fig:pp-lhc}): a steeper fall-down
 of the distribution at large $\eta$. However, the experimental accuracy
 does not allow one to discriminate between the different values of  
 $\alpha_{\rm sea}$.

\begin{figure}[htb]
\centering
\includegraphics[height=6cm,width=0.49\textwidth]{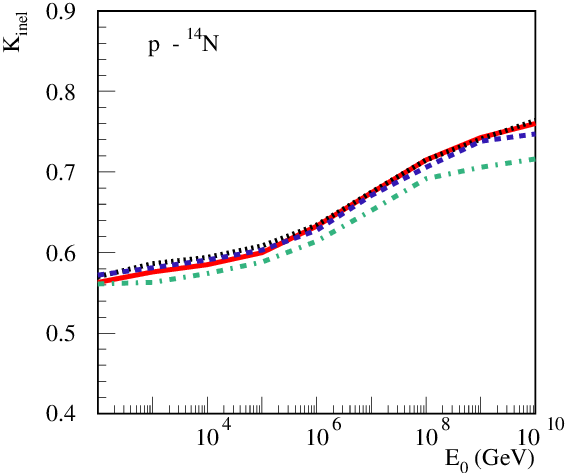}
\caption{Lab.\ energy dependence  of  the inelasticity of
  $p\,^{14}$N collisions, calculated with the default QGSJET-III model
(red solid line) and   with  the option characterized by twice stronger HT corrections, for the parameter  $\alpha_{\rm sea}=0.65$, 0.8, and 0.9  -- 
black dotted, blue dashed, and green dash-dotted lines, respectively.}
\label{fig:kinel-htp}       
\end{figure}

\begin{figure}[htb]
\centering
\includegraphics[height=6cm,width=0.49\textwidth]{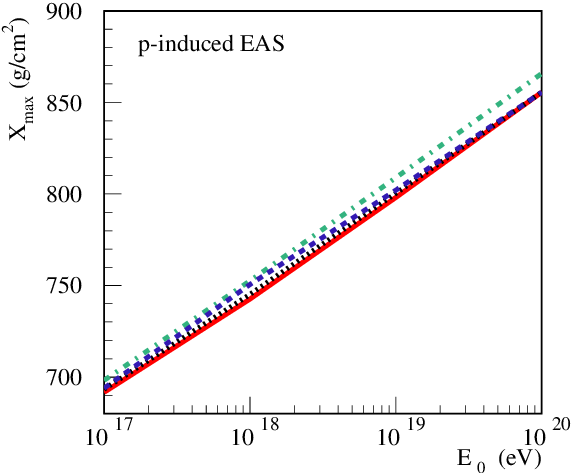}
\caption{Dependence on primary energy of  the average maximum depth 
of proton-initiated EAS.
The notations for the lines are the same as in Fig.\   \ref{fig:kinel-htp}.}
\label{fig:xmax-htp}       
\end{figure}

The impact of the considered modifications of the interaction treatment
on the energy dependence of the inelasticity $K_{\rm inel}$ of proton-nitrogen
collisions is shown in Fig.\ \ref{fig:kinel-htp},
 while the corresponding changes of the predicted
shower maximum depth are plotted in Fig.\  \ref{fig:xmax-htp}. 
As one can see in Figs.\  \ref{fig:kinel-htp} and  \ref{fig:xmax-htp}, 
increasing the strength of HT effects, without modifying momentum distributions
of constituent partons (the case $\alpha_{\rm sea}=0.65$, shown by the black
dotted 
lines in the Figures), does not produce any significant differences, compared
to the default QGSJET-III predictions. This is because such HT corrections
mostly affect relatively central (small impact parameter $b$) collisions
involving high parton densities and being characterized by high multiple
scattering rates, hence, by a very high inelasticity. The overall reduction of
multiple scattering due to stronger HT effects does not have an appreciable
impact on the average  $K_{\rm inel}$ which is rather dominated by
contributions of more peripheral (larger $b$) proton-nucleus collisions.

\begin{figure}[htb]
\centering
\includegraphics[height=4.5cm,width=0.49\textwidth]{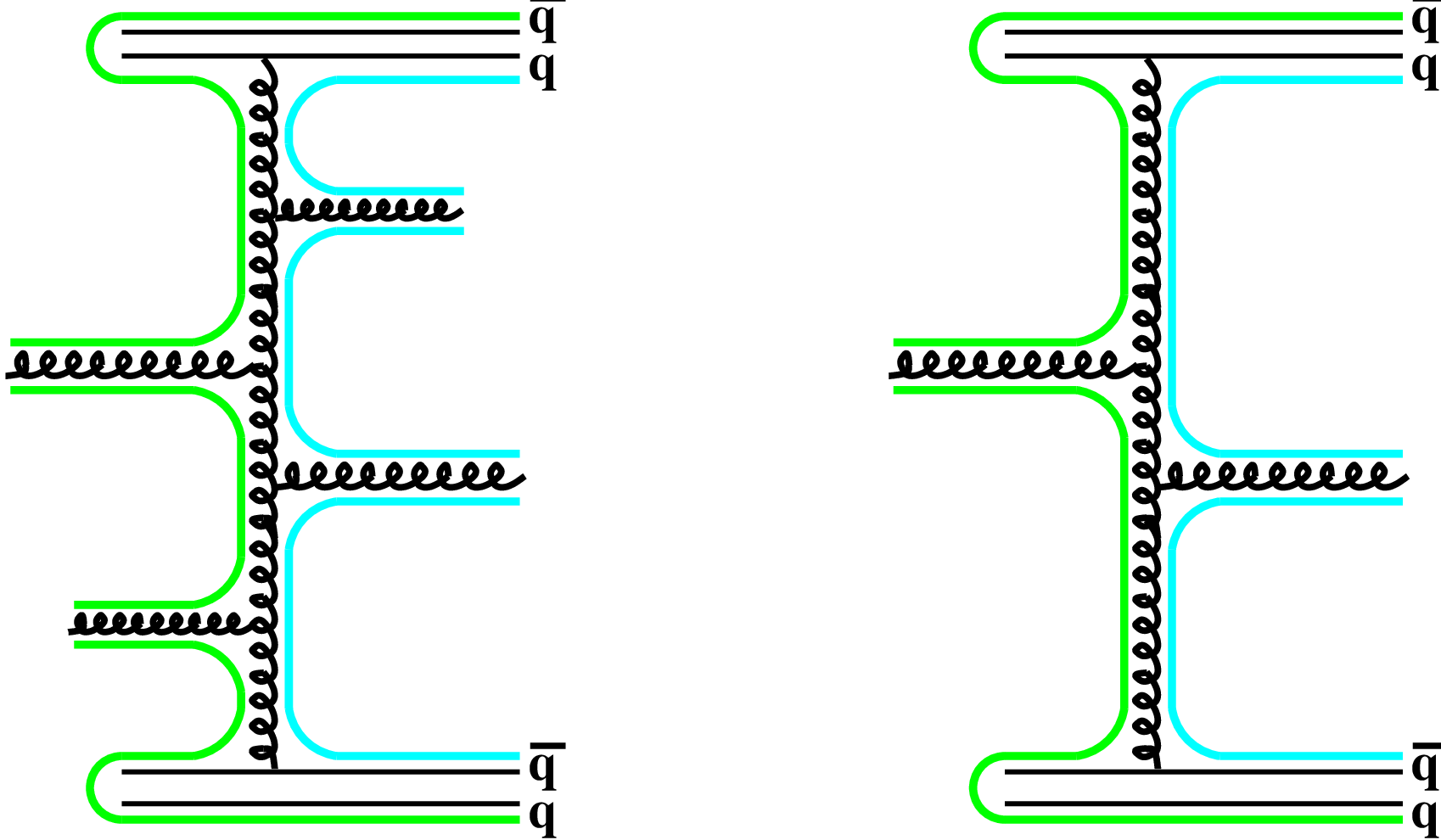}
\caption{Schematic view of a single semihard scattering process. In the standard treatment (left),
strings are formed between constituent partons (quarks and antiquarks) and/or
partons produced by perturbative cascades, following the  color and anticolor
 flows (thick green and blue  lines). Neglecting  ISR (right), strings 
are formed between constituent partons and partons emerging from the hardest scattering process.}
\label{fig:isr-strings}       
\end{figure}

The situation changes noticeably when choosing softer LC momentum distributions
of constituent partons (using $\alpha_{\rm sea}=0.8$ and $\alpha_{\rm sea}=0.9$),
to which strings of color field are connected [see Fig.\ \ref{fig:isr-strings} (left)]. 
The larger
 $\alpha_{\rm sea}$ value is used, the shorter such strings are and the smaller
 fraction of LC momentum of incident proton goes into multiple hadron production 
 resulting from the fragmentation of such strings. As one can see  in
  Figs.\  \ref{fig:kinel-htp} and  \ref{fig:xmax-htp},
 for  $\alpha_{\rm sea}=0.9$, one has up to $\simeq 6$\% reduction of 
  $K_{\rm inel}$ and up to $\simeq 12$ g/cm$^2$ larger $\langle X_{\max}\rangle$
  at the highest energies.
  
  While the considered modifications remain compatible with LHC data
   on secondary hadron production, as demonstrated in Fig.\   \ref{fig:pp-lhc},
   one may wonder why their impact on the 
energy dependence of  $K_{\rm inel}$ and $\langle X_{\max}\rangle$ is so moderate.
This is quite nontrivial, being related to the so-called initial state radiation (ISR)
of partons in hard scattering processes. As discussed, e.g.\ in \cite{ost23},
in any partial semihard rescattering, the hardest (highest $p_t$) parton scattering 
process is typically preceded by multiple emission of ``softer'' partons 
characterized by smaller transverse momenta but having
larger fractions of LC momentum
taken from the parent hadron. In fact, it is this ISR, i.e., 
the perturbative parton cascade preceding the hardest scattering process, which produces large collinear
and infrared logarithms compensating the smallness of the strong coupling constant involved
and gives rise to a steep energy rise of (mini)jet production \cite{glr}. 
The crucial thing here is the $p_t$ and $x$ ordering: each previous parton in the
``ladder'' formed by successive parton emissions is typically characterized by a
much smaller transverse momentum $p_t$  than the next one,
while having a much larger LC
momentum fraction $x$. Therefore, the total LC momentum fraction taken by the
 first $s$-channel partons produced in such
perturbative parton cascades constitutes a natural lower bound on the inelasticity,
whatever soft distribution for nonperturbative constituent partons is chosen \cite{ost23}.

Thus, we arrived here to the point that potential variations of $K_{\rm inel}$
and the corresponding changes of  $X_{\max}$ are restricted by theoretical arguments,
rather than by experimental data. A natural question is how robust are these restrictions.
In principle, one can not  exclude the possibility that the hadron production 
pattern corresponding to the above-discussed picture is modified at sufficiently high 
energies by collective effects. In particular, one popular option is to consider a
rearrangement of ``color flows'', which gives rise to different string configurations,
typically reducing both the total string length in the rapidity space and the fractions
of LC momenta taken by string end partons from the interacting hadrons
 (nuclei)\footnote{An alternative treatment of collective effects,
which involves a thermodynamical description of the hadronization process,
has been discussed, e.g., in \cite{wer24}.} (e.g.\  \cite{chr15}), thereby overcoming the above-discussed lower bound 
  on $K_{\rm inel}$.

Here we prefer to restrain from  relying on a particular treatment of collective effects
in secondary hadron production, rather using an effective approach which allows for
extreme modifications of the predicted inelasticity. Namely, for each semihard rescattering,
we generate the hardest scattering process according to the standard 
collinear factorization formalism of the perturbative quantum chromodynamics,
 while suppressing ISR, such that strings of color field are stretched between 
constituent partons of the interacting hadrons (nuclei) and   final state partons
produced in the hardest process, as sketched in Fig.\ \ref{fig:isr-strings} (right).
 In such a scheme, 
varying the $\alpha_{\rm sea}$ parameter corresponds effectively both to a modification
of momentum distributions of constituent partons and to a varying strength of collective
effects in the hadronization procedure. In particular, in the limit 
$\alpha_{\rm sea}\rightarrow 1$, one ends up with very short strings of color field,
concentrated in the central rapidity region in c.m.\ frame, such that multiple
semihard rescattering processes have a 
minor impact on  $K_{\rm inel}$.
However, it is important to remark that such a limit is nonphysical for two reasons. 
First, potential collective effects may be efficient in relatively central collisions characterized by high parton densities, while being rather weak in more peripheral collisions dominating the average inelasticity. Secondly, the chance that collective effects, however strong they are, eliminate all the partons from ISR should be vanishingly small.

\begin{figure*}[t]
\centering
\includegraphics[height=6.cm,width=\textwidth]{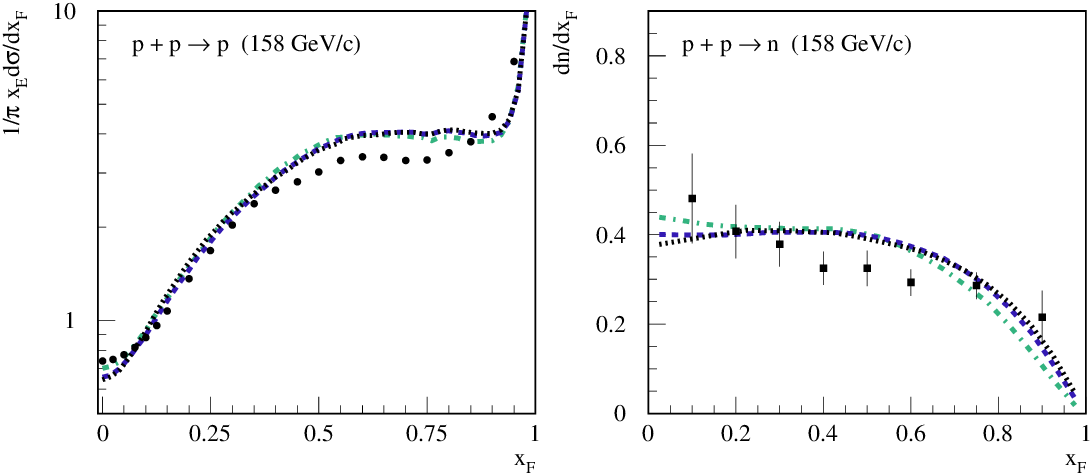}
\caption{$x_{\rm F}$-dependence of $p_t$-integrated invariant cross 
section 
for proton production (left)
and $x_{\rm F}$-distribution of neutrons (right) in c.m.\ frame,
for  $pp$  collisions at 158 GeV/c, compared to NA49 data \cite{ant10}.
Black dotted, blue dashed, and green dash-dotted lines correspond to calculations 
with the modified QGSJET-III model: suppressing ISR and using 
 $\alpha_{\rm sea}=0.65$, 0.8, and 0.9, respectively.}
\label{fig:pp158-evol}       
\end{figure*}

\begin{figure*}[t]
\centering
\includegraphics[height=6.cm,width=0.49\textwidth]{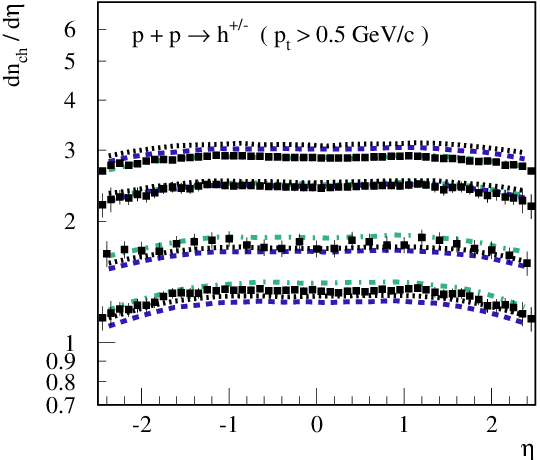}\hfill
\includegraphics[height=6.cm,width=0.49\textwidth]{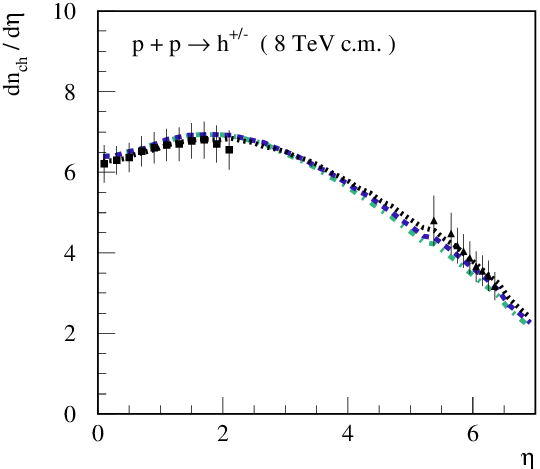}
\caption{Calculated pseudorapidity $\eta$ distributions of
charged hadrons in c.m.\ frame. Left: for $pp$ collisions at 
different $\sqrt{s}$ (from top to bottom: 13, 7, 2.36, and 0.9 TeV),
for hadron transverse momentum $p_t>0.5$  GeV/c, compared to ATLAS 
data \cite{aad11,aad16}. Right:  for $pp$ collisions at $\sqrt{s}=8$
TeV, compared to the data of CMS and TOTEM \cite{cha14}.
The notations for the lines are the same as in Fig.\ \ref{fig:pp158-evol}.}
\label{fig:pp-lhc-evol}       
\end{figure*}

\begin{figure}[htb]
\centering
\includegraphics[height=6cm,width=0.49\textwidth]{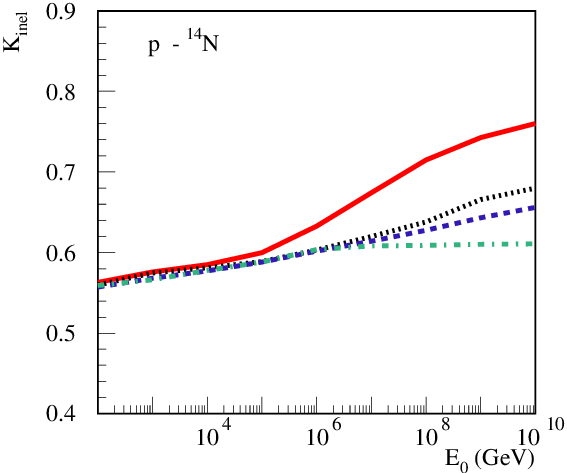}
\caption{Lab.\ energy dependence  of  the inelasticity of
  $p\,^{14}$N collisions, calculated with the default QGSJET-III
(red solid line) and with the modified  model: suppressing ISR and using 
 $\alpha_{\rm sea}=0.65$, 0.8, and 0.9  -- 
black dotted, blue dashed, and green dash-dotted lines, respectively.}
\label{fig:kinel-evol}       
\end{figure}

\begin{figure}[htb]
\centering
\includegraphics[height=6cm,width=0.49\textwidth]{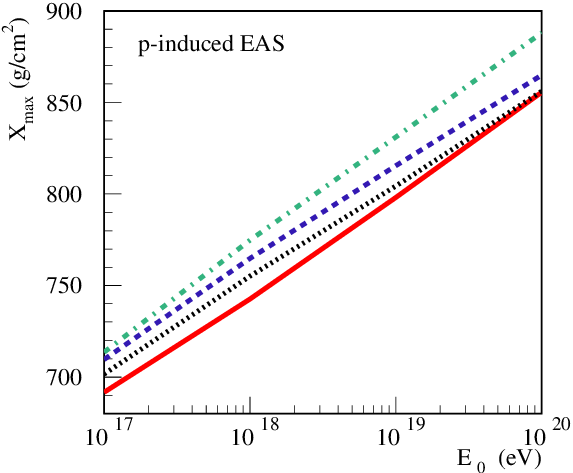}
\caption{Dependence on primary energy of  the average maximum depth 
of proton-initiated EAS. The notations for the 
lines are the same as in Fig.\   \ref{fig:kinel-evol}.}
\label{fig:xmax-evol}       
\end{figure}

As previously, we perform a modeling of hadron-proton and hadron-nucleus interactions,
using different values of $\alpha_{\rm sea}$ and  adjusting other parameters of the 
hadronization procedure  in order to keep an agreement with accelerator measurements,
a comparison with selected  data sets being plotted in 
 Figs.\ \ref{fig:pp158-evol} and \ref{fig:pp-lhc-evol}. 
 As we can see in Fig.\ \ref{fig:kinel-evol}, 
even using the default value $\alpha_{\rm sea}=0.65$, the inelasticity
is reduced by up to $\simeq 10$\% at the highest energies, compared to the QGSJET-III
predictions, which demonstrates the importance of perturbative parton cascades both for hadron production in general and for the energy loss of leading nucleons in particular. 
On the other hand, for 
 $\alpha_{\rm sea}=0.9$,   $K_{\rm inel}$ is practically energy-independent
 above 1 PeV,  where secondary hadron production is dominated by semihard processes.
 The corresponding energy dependence of the predicted $\langle X_{\max}\rangle$,
  for  $\alpha_{\rm sea}=0.65$, 0.8,
 and 0.9, is plotted in Fig.\ \ref{fig:xmax-evol}. 
The reduction of the inelasticity leads to a
 noticeably larger elongation rate for proton-induced EAS; for
  $\alpha_{\rm sea}=0.9$, the shower maximum is $\simeq 30$ g/cm$^2$ deeper at $E_0=10^{20}$ eV, 
  compared to the  QGSJET-III predictions.
  
\begin{figure*}[t]
\centering
\includegraphics[height=11.cm,width=\textwidth]{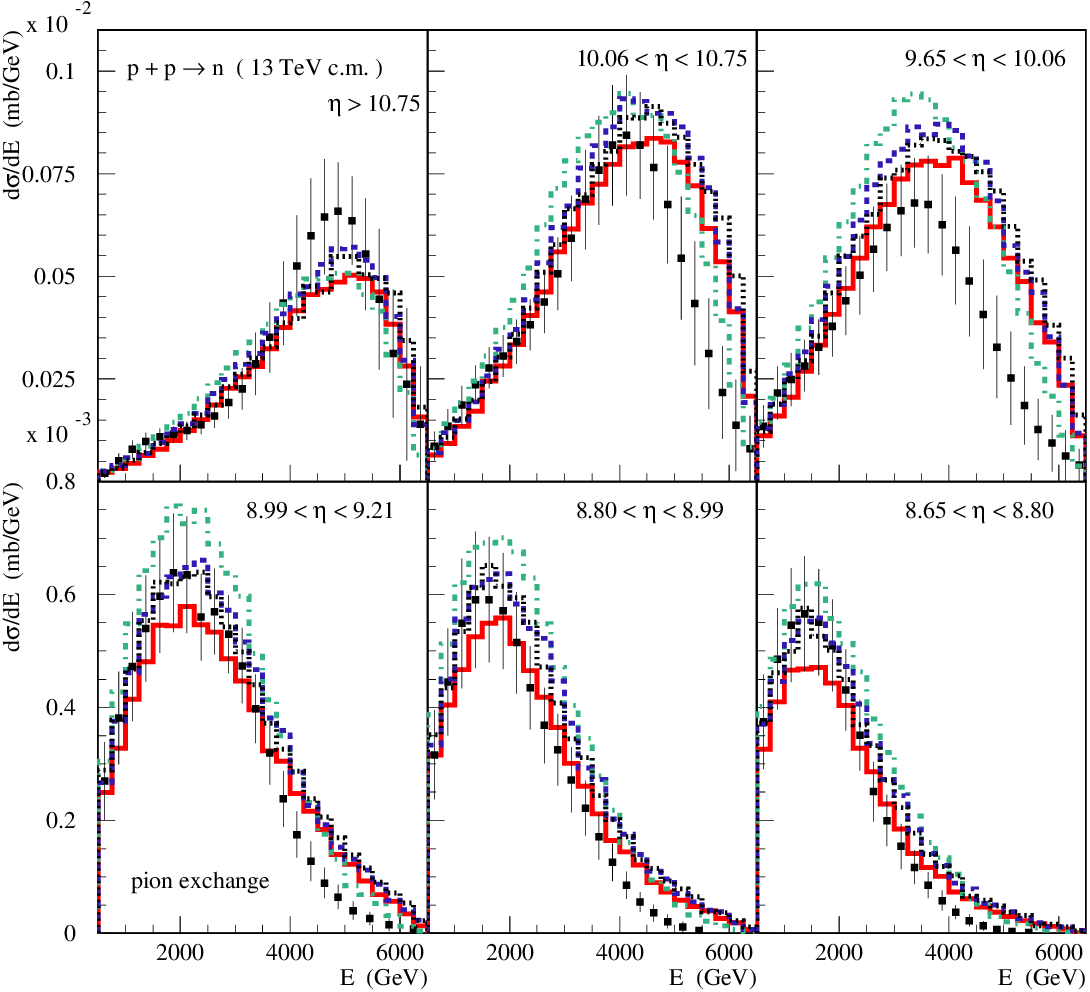}
\caption{Calculated neutron energy spectra  in c.m.\ frame,
 for  $pp$ collisions at $\sqrt{s}=13$ TeV, 
 compared to the data  of the LHCf experiment  \cite{adr18} (points).
 Shown as red solid histograms are the results of the default  QGSJET-III,
 while back dotted, blue dashed, and green dash-dotted histograms correspond to calculations 
 with the modified  model: suppressing ISR and using 
  $\alpha_{\rm sea}=0.65$, 0.8, and 0.9, respectively.}
\label{fig:nlhcf13}       
\end{figure*}

  Now the crucial question is whether such modifications can be constrained by LHC data on forward hadron production.
In Fig.\  
 \ref{fig:nlhcf13}, 
we compare production spectra of 
neutrons,
calculated with the modified QGSJET-III model: suppressing ISR and using  
$\alpha_{\rm sea}=0.65$, 0.8,  and 0.9, to the corresponding measurements of the
LHCf experiment. 
As one can see in   Fig.\ \ref{fig:nlhcf13}, choosing softer momentum 
 distributions of constituent partons, i.e., using larger  $\alpha_{\rm sea}$, 
 one obtains larger neutron yields at high
  $x_{\rm F}\simeq 2E/\sqrt{s}\gtrsim 0.5$
  than observed by 
 the experiment, thereby underestimating  the inelasticity for  neutron 
 production.\footnote{As discussed in \cite{ost24a}, there is a number of indications,
 also from the data of the LHCf experiment, that the QGSJET-III model 
 underestimates somewhat the inelasticity of $pp$ collisions, 
 which is the main reason for its larger  predicted $\langle X_{\max}\rangle$,
  compared to QGSJET-II-04. The modifications of the interaction treatment,
   considered here, further aggravate the tension with the LHCf data on 
   forward neutron production.} Thus, the most extreme modifications of 
   the model, leading to an
 approximate Feynman scaling in the fragmentation region, are 
 somewhat disfavored  by the LHCf data. 
 It is noteworthy that  more stringent constraints may arise from
 studying correlations between central and forward hadron production \cite{ost16},
 e.g., from the ongoing combined measurements of hadron production in $pp$ 
 collisions by the ATLAS and LHCf experiments \cite{atlas17,kon23}.

\begin{figure}[htb]
\centering
\includegraphics[height=6cm,width=0.49\textwidth]{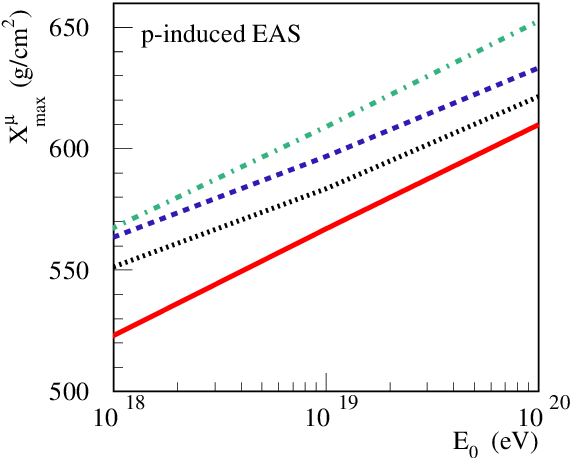}
\caption{Dependence on primary energy of  the  average depth
of maximum of the  muon production profile, 
 for proton-initiated EAS, as calculated with the default 
 QGSJET-III (red solid line) and with the modified  model: suppressing ISR and 
 using $\alpha_{\rm sea}=0.65$, 0.8, and 0.9  -- 
black dotted, blue dashed, and green dash-dotted lines, respectively.}
\label{fig:xmumax-evol}       
\end{figure}

It is worth remarking that the  modifications of the interaction treatment,
 leading to a substantially larger elongation rate, are rather strongly disfavored 
by observations of the  Pierre Auger  Observatory, regarding the 
muon production depth in air showers  \cite{and12,caz12}, notably, by the measured values of
 $\langle X_{\max}^{\mu}\rangle$ -- the average depth of maximum
of the muon production profile  \cite{aab14}.
As discussed in \cite{ost19,ost16a}, 
changes of a model treatment of proton-air collisions, which  produce
 a larger $\langle X_{\max}\rangle$, shift  the average depth of
 maximum of the  muon production profile 
deeper in the atmosphere by a comparable amount. Moreover, since the above-considered
modifications, namely, the suppression of ISR and the use of softer momentum
distributions of constituent partons, impact also pion-air interactions, the 
corresponding effect on  $\langle X_{\max}^{\mu}\rangle$ 
 is stronger, compared to the change of 
 $\langle X_{\max}\rangle$, as demonstrated in Fig.\ \ref{fig:xmumax-evol}.\footnote{Plotted 
 in   Fig.\ \ref{fig:xmumax-evol} is the  the average depth of
 maximum of   muon production profiles produced by EAS simulations, for muon energies
 $E_{\mu}>1$ GeV. This should be distinguished from the so-called apparent muon
 production depth derived by taking into consideration both the impact of muon propagation 
 in the atmosphere and the effects of the corresponding experimental reconstruction
 procedures, regarding measurements with ground-based detectors (see, e.g.\  
  \cite{and12,caz12} for the corresponding discussion).}
For the most extreme modifications,
 the calculated  $\langle X_{\max}^{\mu}\rangle$  appears to be up to
  $\simeq 40$  g/cm$^2$ larger  at the highest energies, than
 predicted by QGSJET-III, thereby creating a strong tension with the Auger
  data.\footnote{While the predictions of the QGSJET-III model for  $\langle X_{\max}^{\mu}\rangle$
  are quite similar to the ones of QGSJET-II-04 \cite{ost24a}, the latter are only consistent
  with the data of the  Pierre Auger  Observatory if one assumes a pure iron 
  composition  for UHECRs \cite{aab14}. The larger $\langle X_{\max}^{\mu}\rangle$ 
  values resulting from the
  considered modifications of the model would thus give rise to a nonphysical
  result: requiring UHECRs to be heavier than iron.} 
 
\section{Summary\label{summary.sec}}
We performed a quantitative analysis of model uncertainties
for predicted maximum depth 
of proton-initiated extensive air showers,
in the framework of the QGSJET-III hadronic interaction model,
restricting ourselves to  the standard physics picture.
Using the conventional approach to the treatment of high energy  interactions,
we investigated a possibility to obtain larger values of 
 $\langle X_{\max}\rangle$ , considering
variations of the inelastic proton-proton cross section, the rate of inelastic 
diffraction, the strength of nonlinear interaction effects, and momentum distributions
of constituent partons involved in multiple scattering processes, allowed by LHC data.
 The studied modifications of the interaction treatment allowed 
us to increase   the predicted  $\langle X_{\max}\rangle$ by only $\sim 10$ g/cm$^2$. 

Such a small variation of the predicted EAS maximum depth,
when modifying  $\sigma^{\rm inel}_{pp}$ and  $\sigma^{\rm diffr}_{pp}$ within the
range allowed by accelerator data, comes at no surprise, given extensive and precise
measurements of proton-proton interaction cross sections at LHC. Yet one could have expected
larger changes of the calculated   $\langle X_{\max}\rangle$
 to result from the other options studied,
because of their potentially strong impact on the inelasticity of proton-proton and
proton-nucleus collisions. However, potential variations of the inelasticity appeared
to be limited by the initial state radiation of partons in semihard scattering processes:
the total fraction of the incident hadron momentum, taken by all such perturbatively
generated partons constitutes a natural lower bound on the inelasticity.

We further investigated a more exotic scenario, considering a potentially significant 
modification of the parton hadronization procedure by hypothetical collective effects.
That way, we were able to change drastically the predicted energy dependence of the 
inelasticity of proton-air collisions and to increase thereby the predicted EAS maximum 
depth by up to $\simeq 30$   g/cm$^2$. However, those most extreme modifications 
appeared to be disfavored both by the data of the LHCf experiment, regarding forward 
neutron production in $pp$ collisions at LHC,
and by measurements of the  muon production depth by the Pierre Auger Observatory.

\subsection*{Acknowledgments}
 The work of S.O.\ was supported by  Deutsche Forschungsgemeinschaft 
(project number 465275045).  G.S.\ acknowledges
support by the Bundesministerium f\"ur Bildung
und Forschung, under grants 05A20GU2 and 05A23GU3.


\begin{thebibliography}{99}

\bibitem{nag00}
M.\ Nagano  and A.\ A.\ Watson, 
{\em Observations and implications of the ultrahigh-energy cosmic rays},
Rev.\ Mod.\ Phys.\ {\bf 72}, 689 (2000).

\bibitem{hec98}
D.\ Heck, J.\ Knapp, J.\ N.\ Capdevielle, G.\ Schatz, and T.\ Thouw, 
{\em CORSIKA: A Monte Carlo
code to simulate extensive air showers},
 Forschungszentrum Karlsruhe Internal Report FZKA-6019 (1998).

\bibitem{eng11} 
R.\ Engel, D.\ Heck, and T.\ Pierog, 
{\em Extensive air showers and hadronic interactions at high energy},
Ann.\ Rev.\ Nucl.\ Part.\ Sci.\  {\bf 61}, 467 (2011).

\bibitem{kam12} 
	 K.-H. Kampert and M. Unger, {\em Measurements of the Cosmic Ray 
	 Composition with Air Shower Experiments},
 Astropart.\ Phys.\ {\bf 35}, 660 (2012).

\bibitem{ost19} 
 S.\ Ostapchenko, 
 {\em High energy interactions of cosmic rays},
   Adv.\ Space Res.\   {\bf  64},  2445  (2019).  
 
\bibitem{abr13}
P.\ Abreu {\em et al.}  (Pierre Auger Collaboration), 
{\em Interpretation of the Depths of Maximum of
Extensive Air Showers Measured by the Pierre Auger Observatory}, 
JCAP {\bf 02}, 026 (2013).

\bibitem{PierreAuger:2024neu}
A.~Abdul Halim {\em et al.} (Pierre Auger Collaboration),
{\em Testing Hadronic-Model Predictions of Depth of 
Maximum of Air-Shower Profiles and Ground-Particle Signals 
using Hybrid Data of the Pierre Auger Observatory},
 Phys.\ Rev.\  D \textbf{109},  102001  (2024).  

\bibitem{pao14}  
A.\ Aab  {\em et al.} (Pierre Auger Collaboration),
{\em Depth of maximum of air-shower profiles at the Pierre Auger
 Observatory. I. Measurements at energies above $10^{17.8}$ eV},
 Phys.\ Rev.\  D \textbf{90}, 122005 (2014).  

\bibitem{pao14a}  
A.\ Aab   {\em et al.} (Pierre Auger Collaboration),
{\em Depth of maximum of air-shower profiles at the Pierre Auger 
Observatory. II. Composition implications},
 Phys.\ Rev.\  D \textbf{90},  122006 (2014).  

\bibitem{ost11} S. Ostapchenko, 
 {\em Monte Carlo treatment of hadronic 
interactions in enhanced Pomeron scheme: QGSJET-II model}, 
 Phys.\  Rev.\ D {\bf 83},  014018 (2011).

  \bibitem{ost13}
S.~Ostapchenko,
  {\em QGSJET-II: physics, recent improvements, and results for
 air showers},
  EPJ Web Conf.\   {\bf 52},    02001 (2013).
 
   \bibitem{ost24} S.\ Ostapchenko, 
 {\em QGSJET-III model of high energy hadronic interactions: The formalism},
 Phys.\  Rev.\ D {\bf  109},   034002 (2024).  
 
   \bibitem{ost24a} S.\ Ostapchenko, 
 {\em QGSJET-III model of high energy hadronic interactions: II. Particle
production and extensive air shower characteristics},
 Phys.\  Rev.\ D {\bf  109},   094019  (2024).  
 
\bibitem{nmu24}
S.\ Ostapchenko and G.\ Sigl, 
 {\em On the model uncertainties for the predicted muon
content of extensive air showers}, 
 Astropart.\ Phys.\ {\bf 163},  103004 (2024).

\bibitem{ant19}
G.\ Antchev  {\em et al.} (TOTEM Collaboration),
 {\em   First measurement of elastic, inelastic and total cross-section at
   $\sqrt{s}=13$ TeV by TOTEM
and overview of cross-section data at LHC energies},
 Eur.~Phys.~J.~C {\bf 79}, 103  (2019).

\bibitem{aad23}
G.\ Aad   {\em et al.} (ATLAS Collaboration),
 {\em  Measurement of the total cross section and $\rho$-parameter from elastic scattering in pp collisions at $\sqrt{s}=13$ TeV with the ATLAS detector},
 Eur.~Phys.~J.~C {\bf 83},  441  (2023).
 
\bibitem{gla59}
R.\ J.\ Glauber, 
 {\em High-energy collision theory},
in: Lectures in theoretical physics, 
Ed.\ by W.\ E.\ Brittin and
L.\ G.\ Dunham, Interscience Publishers, New York,
1959, vol. 1, p. 315.

\bibitem{gri69}
	 V.\ N.\ Gribov, 
	  {\em Glauber corrections and
the interaction between high-energy hadrons
and nuclei}, 
Sov.\ Phys.\ JETP {\bf 29}, 483 (1969).

\bibitem{ost16}
S.\ Ostapchenko, M.\ Bleicher, T.\ Pierog, and K.\ Werner, 
{\em Constraining high energy interaction
mechanisms by studying forward hadron production
at the LHC}, 
Phys.\ Rev.\  D    {\bf 94},   114026 (2016).

 \bibitem{alo08}
R. Aloisio, V. Berezinsky, P. Blasi, and S.~Ostapchenko, 
{\em Signatures of the transition from
Galactic to extragalactic cosmic rays}, 
Phys.\ Rev.\  D    {\bf 77},    025007 (2008).

\bibitem{ant13b}
G.\ Antchev   {\em et al.} (TOTEM Collaboration),
 {\em   Luminosity-independent measurements of total, elastic and inelastic cross-sections at $\sqrt{s}=7$ TeV},
Europhys.\ Lett.\  {\bf 101}, 21004 (2013).

\bibitem{ant13a}
G.\ Antchev  {\em et al.} (TOTEM Collaboration),
 {\em  Measurement of proton-proton inelastic scattering cross-section at  $\sqrt{s}=7$ TeV},
Europhys.\ Lett.\  {\bf 101}, 21003 (2013).

\bibitem{ant13c}
G.\ Antchev  {\em et al.} (TOTEM Collaboration),
 {\em   Luminosity-Independent Measurement of the Proton-Proton Total Cross Section at  $\sqrt{s}=8$ TeV},
Phys.\ Rev.\  Lett.\    {\bf 111}, 012001 (2013).
   
\bibitem{aad14}
G.\ Aad  {\em et al.} (ATLAS Collaboration),
 {\em  Measurement of the total cross section from elastic scattering in $pp$
collisions at $\sqrt{s}=7$ TeV with the ATLAS detector},
 Nucl.~Phys.~B~{\bf 889}, 486 (2014).

\bibitem{aab16}
M.\  Aaboud  {\em et al.} (ATLAS Collaboration),
 {\em  Measurement of the total cross section from elastic scattering in
 $pp$ collisions at  $\sqrt{s}=8$ TeV with the ATLAS detector},
Phys.\ Lett.\ B  {\bf 761},   158 (2016).

\bibitem{ant13}
G.\ Antchev  {\em et al.} (TOTEM Collaboration),
 {\em  Double diffractive cross-section measurement in the forward 
 region at the LHC}, 
Phys.\ Rev.\  Lett.\    {\bf 111},  262001 (2013).
	
\bibitem{aad22}
G.\ Aad   {\em et al.} (ATLAS Collaboration),
 {\em Measurement of differential cross sections for single
diffractive dissociation in $\sqrt{s}=8$ TeV $pp$ collisions
using the ATLAS ALFA spectrometer},
 J.\  High Energy Phys.\  {\bf 02},  042 (2020).

 \bibitem{olj20}  
 F.~Oljemark,
{\em Single Diffraction in proton-proton scattering with TOTEM 
at the Large Hadron Collider}, PhD thesis,
University of Helsinki (2020).

\bibitem{ost14}
 S.\ Ostapchenko,
 {\em  LHC data on inelastic diffraction
and uncertainties in the predictions for longitudinal
extensive air shower development}, 
Phys.\ Rev.\  D    {\bf 89},  074009 (2014).

\bibitem{eng92}
J.\ Engel, T.\ K.\ Gaisser, T.\ Stanev, and P.~Lipari,
{\em Nucleus-nucleus collisions and interpretation of 
cosmic ray cascades}, 
Phys.\  Rev.\ D {\bf 46}, 5013 (1992).

\bibitem{kal93}
N.\ N.\ Kalmykov and S.\ S.\ Ostapchenko, 
{\em The nucleus-nucleus interaction, nuclear fragmentation,
and fluctuations of extensive air showers}, 
Phys.\ Atom.\ Nucl.\ {\bf 56}, 346 (1993).

\bibitem{bia76}
A.\ Bia{\l}as, M.\  Bleszynski, and W.\  Czyz, 
{\em Multiplicity Distributions in Nucleus-Nucleus Collisions at High-Energies},
Nucl.\ Phys.\ B  {\bf  111}, 461 (1976).

\bibitem{kal89}
N.\ N.\ Kalmykov and S.\ S.\ Ostapchenko, 
{\em Comparison of Nucleus-Nucleus Interaction Characteristics
 in the Model of Quark-Gluon Strings and in the Superposition Model},
        Sov.\ J.\ Nucl.\ Phys.\  {\bf 50}, 315 (1989).
	
\bibitem{fre87}
  S.\ Fredriksson, G.\ Eilam, G.\ Berlad, and L.\ Bergstr\"om, 
{\em High-energy Collisions With Atomic Nuclei. Part 1},
 Phys.\ Rept.\  {\bf 144}, 187 (1987).
 
\bibitem{goo60} 
M.~L.~Good and W.~D.~Walker,  
 {\em Diffraction disssociation of beam particles},
 Phys.~Rev.~{\bf 120}, 1857  (1960).
   
\bibitem{kho21} V.~A.~Khoze, A.~D.~Martin, and M.~G.~Ryskin,
  {\em Dynamics of diffractive dissociation},
Eur.~Phys.~J.~C~{\bf 81},  175  (2021).

  \bibitem{ost19a}
S.~Ostapchenko and   M.~Bleicher, 
 {\em Taming the energy rise of the total proton-proton cross-section},
Universe  {\bf 5},   106 (2019).

	\bibitem{pdg}
 R.\ L.\ Workman  {\em et al.} (Particle Data Group), 
 {\em Review of Particle Physics},
   Prog.\ Theor.\ Exp.\ Phys.\  {\bf 2022}, 083C01 (2022).
 
\bibitem{ber07}
T.\ Bergmann, R.\ Engel, D.\ Heck, N.~N.~Kalmykov, S.~Ostapchenko,
 T.~Pierog, T.~Thouw, and K.\ Werner,
 {\em  One-dimensional Hybrid Approach to Extensive Air Shower Simulation},
  Astropart.\ Phys.\   {\bf 26}, 420 (2007).

\bibitem{ost03}
S.\ S.\ Ostapchenko,
 {\em Contemporary models of high-energy interactions:
  Present status and perspectives},
 J.\ Phys.\ G {\bf 29}, 831 (2003).

\bibitem{par11}
  R.\ D.\ Parsons, C.\  Bleve, S.\ S.\  Ostapchenko, and J.\  Knapp,
 {\em   Systematic uncertainties in air shower measurements from 
 high-energy hadronic interaction models},
  Astropart.\ Phys.\   {\bf 34},  832 (2011).

  \bibitem{aar10}
 F.~D.~Aaron    {\em et al.} (H1 and ZEUS Collaborations),
 {\em Combined measurement and QCD analysis of the inclusive $e^{\pm}p$ scattering cross sections at HERA},
 J.\  High Energy Phys.\  {\bf 01}, 109 (2010).
   
	\bibitem{ant10}
   T.\ Anticic  {\em et al.} (NA49 Collaboration), 
  {\em   Inclusive production of protons, anti-protons
and neutrons in p+p collisions at 158-GeV/c beam momentum}, 
Eur.\ Phys.\ J.\ C   {\bf 65}, 9 (2010).

	\bibitem{aad11}
 G.\ Aad  {\em et al.} (ATLAS Collaboration),
  {\em Charged-particle multiplicities in pp interactions measured with 
  the ATLAS detector at the LHC}, 
  New J.\ Phys.\  {\bf 13}, 053033 (2011).
  
	\bibitem{aad16}
 G.\ Aad  {\em et al.} (ATLAS Collaboration),
  {\em   Charged-particle distributions in $\sqrt{s}=13$
TeV pp interactions measured with the ATLAS detector at the LHC}, 
Phys.\ Lett.\ B  {\bf 758}, 67 (2016).

	\bibitem{cha14}
S.\ Chatrchyan  {\em et al.} (CMS and TOTEM Collaborations), 
 {\em Measurement of pseudorapidity distributions of charged particles 
 in proton-proton collisions at $\sqrt{s}=8$ TeV by
the CMS and TOTEM experiments}, 
Eur.\ Phys.\ J.\ C  {\bf 74}, 3053 (2014).

 \bibitem{ost23}
S.~Ostapchenko, 
{\em Cosmic ray interactions in the atmosphere: QGSJET-III and other models},
SciPost Phys.\ Proc.\  {\bf 13}, 004 (2023).

 \bibitem{glr}
L.\ V.\ Gribov, E.\ M.\ Levin, and M.~G.~Ryskin,
{\em Semihard Processes in QCD},
    Phys.\ Rept.\  {\bf 100},  1 (1983).
 
\bibitem{wer24}
K.\ Werner, 
{\em Core-corona procedure and microcanonical hadronization to understand strangeness enhancement in proton-proton and heavy ion collisions in the EPOS4 framework}, 
Phys.\ Rev.\  C    {\bf  109},   014910 (2024).

 \bibitem{chr15}
 J.\ R.\ Christiansen and P.\ Z.\ Skands,
{\em String Formation Beyond Leading Colour},
 J.\  High Energy Phys.\   {\bf 08}, 003 (2015).
  
 
  \bibitem{adr18} 
 O.\ Adriani   {\em et al.} (LHCf Collaboration),
{\em Measurement of inclusive forward neutron production cross 
 section in proton-proton collisions at $\sqrt{s}=13$ TeV with
  the LHCf Arm2 detector},
  J.\  High Energy Phys.\    {\bf 11}, 073 (2018). 
  
  \bibitem{atlas17} ATLAS and LHCf Collaborations, 
{\em   Measurement of contributions of diffractive
processes to forward photon spectra in pp collisions at $\sqrt{s}=13$  TeV},
 Tech.\ Rep.\ ATLAS-CONF-2017-075, CERN, Geneva, 2017.
  
  \bibitem{kon23} 
  M.\ Kondo   {\em et al.} (LHCf Collaboration),
{\em Performance evaluation of LHCf-ATLAS ZDC joint measurement using proton beam},
  EPJ Web Conf.\   {\bf 283},   05012   (2023).

 \bibitem{and12}  
S.\ Andringa, L.\ Cazon, R.\ Conceicao, and M.\ Pimenta,
{\em The Muonic longitudinal shower profiles at production},
   Astropart.\ Phys.\   {\bf 35}, 821 (2012).

 \bibitem{caz12}  
L.\ Cazon, R.\ Conceicao, M.\ Pimenta, and E.\ Santos,
{\em A model for the transport of muons in extensive air showers},
   Astropart.\ Phys.\   {\bf 36}, 211 (2012).

 \bibitem{aab14}  
A.\ Aab {\em et al.}  (Pierre Auger Collaboration),
{\em Muons in Air Showers at the Pierre Auger Observatory: 
Measurement of Atmospheric Production Depth},
	 Phys.\ Rev.\  D  {\bf 90}, 012012 (2014).

 \bibitem{ost16a}
S.~Ostapchenko and M.~Bleicher, 
{\em Constraining pion interactions at very high energies 
by cosmic ray data},
Phys.\ Rev.\  D    {\bf 93},   051501(R) (2016).

\end{thebibliography}
\end{document}